\documentclass[prb,twocolumn,amsmath,amssymb]{revtex4-1}

\usepackage{graphicx}
\usepackage{dcolumn}
\usepackage{bm}

\begin{document}


\title{Giant Faraday rotation in single- and multilayer graphene}

\author{Iris Crassee$^{1}$, Julien Levallois$^{1}$, Andrew L. Walter$^{2,3}$,
Markus Ostler$^{4}$, Aaron Bostwick$^{3}$, Eli Rotenberg$^{3}$,
Thomas Seyller$^{4}$, Dirk van der Marel$^{1}$, Alexey B.
Kuzmenko$^{1}$\footnote{To whom correspondence should be
addressed Alexey.Kuzmenko@unige.ch}}

\affiliation{
\\
\mbox{ }
\\
$^{1}$D\'epartement de Physique de la Mati\`ere Condens\'ee,
Universit\'e de Gen\`eve, CH-1211 Gen\`eve 4,
Switzerland\\
\\
$^{2}$Dept. of Molecular Physics, Fritz-Haber-Institut der
Max-Planck-Gesellschaft, Faradayweg 4-6, 14195 Berlin,
Germany\\
\\
 $^{3}$E. O. Lawrence Berkeley Laboratory, Advanced
Light Source, MS6-2100, Berkeley, CA 94720\\
\\
$^{4}$Lehrstuhl f\"{u}r Technische Physik, Universit\"{a}t
Erlangen-N\"{u}rnberg, Erwin-Rommel-Str. 1, 91058 Erlangen,
Germany }

\maketitle

\date{\today}

\textbf{Optical Faraday rotation is one of the most direct and
practically important manifestations of magnetically broken
time-reversal symmetry. The rotation angle is proportional to
the distance traveled by the light, and up to now sizeable
effects were observed only in macroscopically thick samples and
in two-dimensional electron gases with effective thicknesses of
several nanometers. Here we demonstrate that a \emph{single
atomic} layer of carbon - graphene - turns the polarization by
several degrees in modest magnetic fields. The rotation is found to be
strongly enhanced by resonances originating from the
cyclotron effect in the classical regime and the
inter-Landau-level transitions in the quantum regime. Combined
with the possibility of ambipolar doping, this opens pathways
to use graphene in fast tunable ultrathin infrared
magneto-optical devices.}


Graphene's unusual physical
properties\cite{GeimNM07,CastroNetoRMP09} make it highly
attractive for both fundamental research and novel
applications. Importantly, the conduction and valence bands in
graphene show a relativistic massless dispersion and a chiral
character of the electronic wavefunctions, resulting in
non-equidistant Landau levels (LLs) ($n = 0, \pm 1, \pm 2...$):
\begin{equation}
E_n=\mbox{sign}(n) \sqrt{2 e \hbar v_F^2 |nB|} \label{EqEn}
\end{equation}
\noindent in a perpendicular magnetic field $B$, which includes
an anomalous LL at zero energy $E_{0}=0$ ($e>0$ is the
elementary charge and $v_F\approx 10^6$ m/s is the Fermi
velocity). As a consequence, graphene shows the half-integer
quantum Hall effect \cite{GusyninPRL05,PeresPRB06}, first
observed in exfoliated graphene flakes
\cite{NovoselovNat05,ZhangNat05} and later, in graphene
epitaxially grown on SiC \cite{ShenAPL09,WuAPL09,JobstPRB10}.

The optical properties of graphene are equally unusual. The
relativistic and gapless dispersion results in a universal and
constant optical conductivity of $\sigma_0 = e^2/4\hbar$
\cite{AndoJPSJ02,PeresPRB06,FalkovskyEPJ07,KuzmenkoPRL08,NairSc08,LiNaturePhys08,GusyninNJP09}
for the photon energies $\hbar\omega$ above twice the Fermi
energy ($\epsilon_F$). In magnetic field, the non-equidistant
spacing of the LLs gives rise to a spectacular series of
absorption peaks, corresponding to transitions between various
levels
\cite{SadowskiPRL06,JiangPRL07,GusyninJPCM07,OrlitaPRL08}. In
contrast, only one peak, centered at the cyclotron frequency
$\omega_c = eB/m_{c}$, where $m_{c}$ is the cyclotron mass, is
observed in conventional 2D electron gases (2DEGs), produced in
semiconductor heterostructures.

\begin{figure*}
\includegraphics[width=15cm]{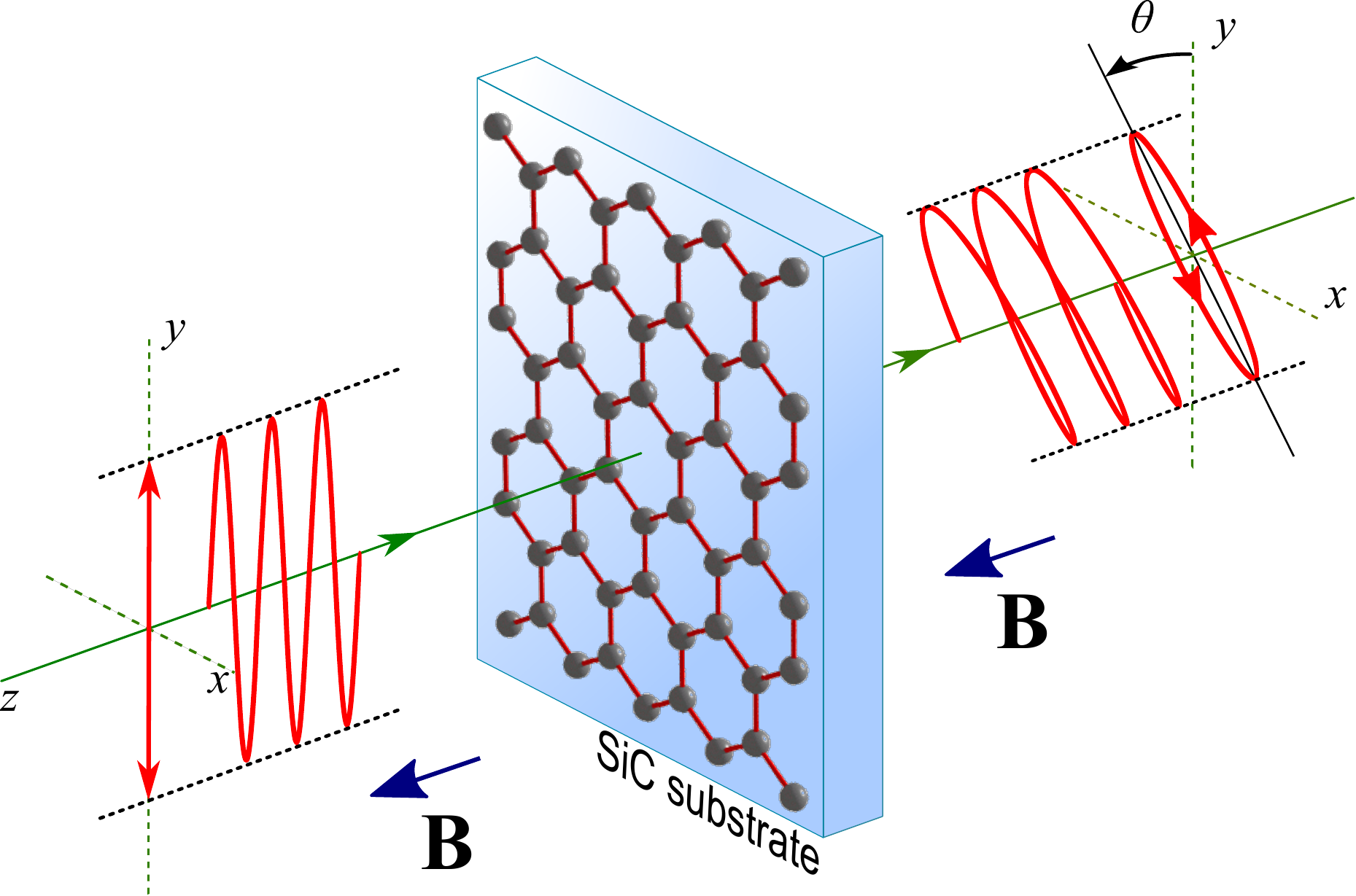}\\
\caption{\textbf{Schematics of the magneto-optical Faraday
rotation experiment}. The polarization plane of the linearly
polarized incoming beam is rotated by the Faraday angle
$\theta$ after passing through graphene on a SiC substrate in a
perpendicular magnetic field. Simultaneously, the polarization
acquires a certain ellipticity. The positive direction of the
magnetic field is along the $z$-axis.} \label{Fig1}
\end{figure*}

Up to now, experimental magneto-optical studies of graphene
\cite{SadowskiPRL06,JiangPRL07,OrlitaPRL08} were focused on the
diagonal conductivity component $\sigma_{xx}(\omega,B)$; the
off-diagonal, or AC Hall\cite{KaplanPRL96}, conductivity
$\sigma_{xy}(\omega,B)$ was addressed however only
theoretically
\cite{GusyninJPCM07,MorimotoPRL09,FialkovskyJPAMT09}. The
diagonal and Hall conductivities  are directly related to the
absorption $1-T$ and Faraday rotation angle $\theta$
(Fig.\ref{Fig1}) respectively. Adopting the thin-film
approximation and keeping only the terms linear in conductivity
\cite{SadowskiPRL06,MorimotoPRL09} the following expressions
are obtained:
\begin{eqnarray}
1-T(\omega,B) &\approx& 2 Z_0 f_s(\omega) \mbox{Re}\left[\sigma_{xx}(\omega,B)\right] \label{EqAbsSxx}\\
\theta(\omega,B) &\approx& Z_0 f_s(\omega)
\mbox{Re}\left[\sigma_{xy}(\omega,B)\right]\label{EqThetaSxy}
\end{eqnarray}
\noindent where $T$ is the substrate-normalized transmission
(see the Methods section),
$Z_0$ $\approx$ 377 $\Omega$ is the impedance of vacuum and
$f_{s}(\omega)$ is a spectrally featureless function specific
to the substrate (as described in the Methods section). Notably, the
diagonal conductivity is formally independent of the sign of
the charge carriers, while the DC Hall conductivity, which is
sensitive to the carrier type, lacks spectral information.
Therefore, studying the Faraday rotation is needed to complete
the picture of the dynamics of electrons and holes in graphene.

Graphene, epitaxially grown on silicon carbide, is well adapted
for magneto-optical studies because of its well-controlled
morphology and essentially unlimited size
\cite{BergerJPCB04,EmtsevNatMater09}. Here, we used single and
multi layer graphene (SLG and MLG), grown respectively on the
Si-terminated \cite{EmtsevNatMater09} and C-terminated
\cite{BergerJPCB04} surfaces of 6H-SiC. The first sample
underwent H-passivation of the Si dangling bonds
\cite{RiedlPRL09,SpeckMSF10}, resulting in quasi-free standing
SLG. The MLG consists of 4 - 6 rotated atomic layers (see the
Methods section). These samples allowed us to explore the cases
of both electron and hole doping as well as to access both the
classical (high doping) and quantum (low doping) regimes as
described below.

\begin{figure*}
\includegraphics[width=17cm]{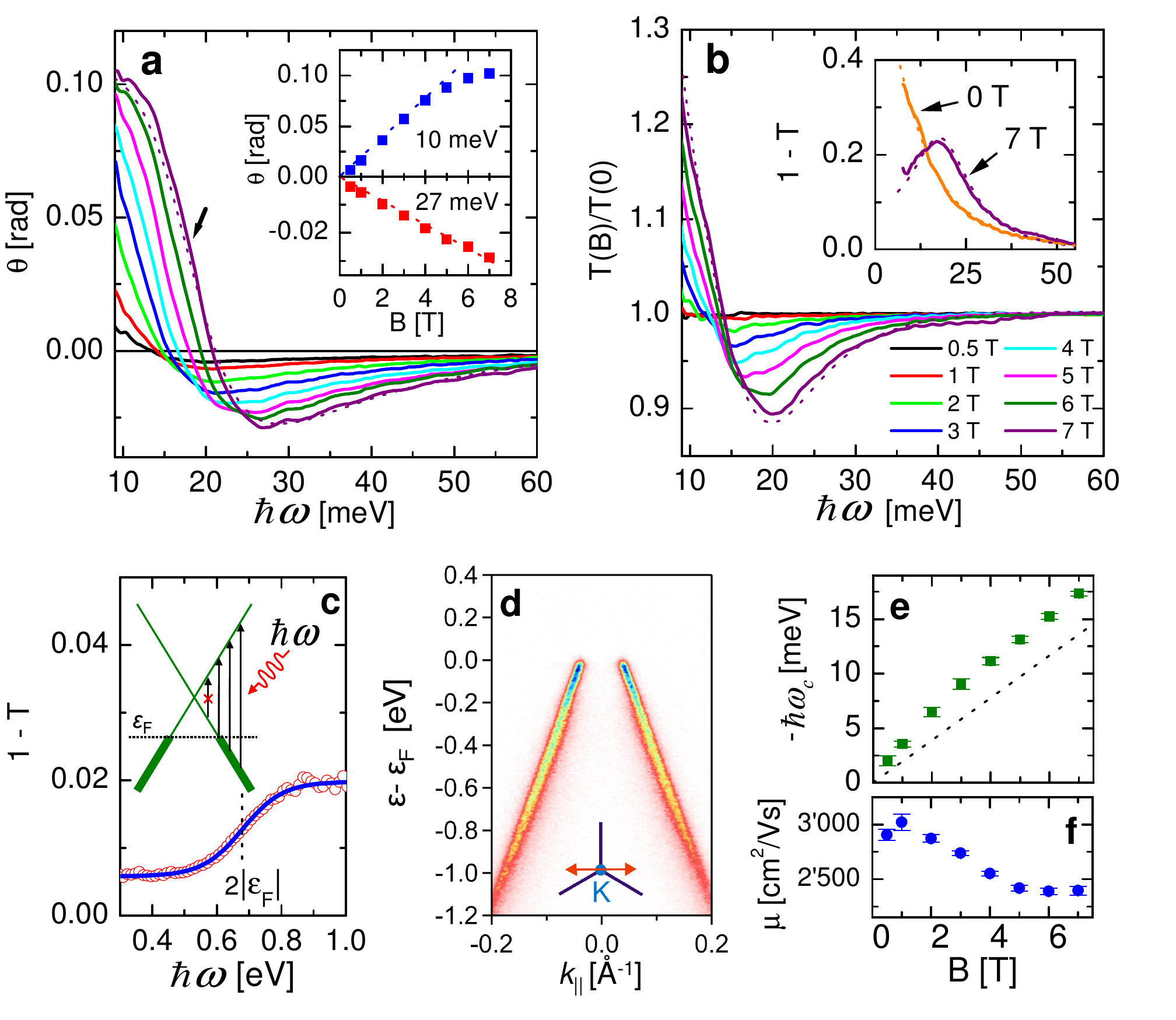}\\
\caption{\textbf{Faraday rotation and magneto-optical
transmission spectra of single layer graphene.} Panel
\textbf{a}: the Faraday angle $\theta$ at several fields up to
7 T at 5 K. The inset presents the magnetic field dependence of
$\theta(B)$ at $\hbar\omega = $10 and 27 meV. The dashed lines
are linear fits of the data points between 0 and 5 T. Panel
\textbf{b}: the transmission ratio $T(B)/T(0)$ at the same
fields. In the inset the absorption spectra $1-T(B)$ for $B$ =
0 T and 7 T are shown. The dashed lines in both panels and the
inset of panel \textbf{b} are fits using the classical
cyclotron resonance equations (\ref{EqSxx}) and (\ref{EqSxy}).
Panel \textbf{c}: the absorption of the same sample at room
temperature in zero field (symbols), showing a clear step at
$\hbar\omega = 2|\epsilon_{F}|$ due to Pauli blocking as shown
in the inset. The blue line is a fitting curve using a
phenomenological broadened step function
$a+b\tanh[(\hbar\omega-2|\epsilon_F|)/\delta\epsilon]$ giving
$|\epsilon_{F}|$ = 0.34 $\pm$ 0.01 eV. Panel \textbf{d}: the
band dispersion near the K-point measured by ARPES on another
sample prepared in the same way. Panel \textbf{e} shows the the
cyclotron energy as a function of $B$. The dashed line is a
theoretical dependence based on equation (\ref{EqWc}) using
$v_F = 1.02\times 10^6$ and $\epsilon_{F}$ = -0.34 eV. Panel
\textbf{f} - the field dependence of the carrier mobility.} \label{Fig2}
\end{figure*}

Figure \ref{Fig2}\textbf{a} shows the Faraday angle $\theta$
measured on SLG at 5 K in magnetic fields up to 7 T in the
far-infrared range as described in the Methods section. The
spectra show a strongly field dependent edge-like structure,
giving rise to a positive rotation at low energies and negative
$\theta$ at high energies. The maximum Faraday rotation exceeds
0.1 radians ($\sim$ 6 degrees), which is an exceptionally large
effect given that it comes from a single layer. Measurements on
the bare substrate did not reveal any Faraday effect, hence the
observed rotation comes exclusively from the carbon monolayer.
The inset shows the field dependence of $\theta$ at 10 and 27
meV. The curves follow an approximately linear dependence, with
opposite slopes of +18.5 mrad/T and -4.5 mrad/T respectively.

The zero-field normalized transmission spectra $T(B)/T(0)$ of
SLG also reveal a strong magnetic field dependence (Fig.
\ref{Fig2}\textbf{b}). The inset shows the absorption $(1-T)$
at 0 and 7 T. One can clearly see a strong Drude peak,
signaling a high doping level. The peak center shifts from zero
to a finite energy at 7 T, which is due to the cyclotron
resonance as we discuss below. Although an estimate of the
doping can be made from the integrated intensity of the Drude
peak, the most direct measurement of the Fermi energy comes
from the absorption spectra in the mid-infrared range (Fig.
\ref{Fig2}\textbf{c}), where an absorption edge is easily
recognized. This is expected to be at 2$|\epsilon_F|$ due to
Pauli blocking
\cite{AndoJPSJ02,PeresPRB06,FalkovskyEPJ07,LiNaturePhys08,GusyninNJP09},
giving an estimate for $|\epsilon_F|$ = 0.34 $\pm$ 0.01 eV.

Independently, we have performed angle resolved photoemission
spectroscopy (ARPES) on a similar sample (Fig.
\ref{Fig2}\textbf{d}). The obtained dispersion curves show that
the sample is p-doped (the Fermi level is in the valence band).
The Fermi energy inferred from optical measurements is in a
reasonable agreement with an extrapolation of the occupied
bands measured by ARPES. Below we show that the carrier type
can also be directly extracted from the Faraday rotation.

The doping level and field strength used here puts the system
in the classical regime, where the separation between the LLs
at the Fermi energy is much smaller than $\epsilon_F$
\cite{OrlitaPRL08,GusyninNJP09}. In this case, the Dirac
quasiparticles are expected to exhibit the classical cyclotron
resonance effect. Indeed, the data in Figures
\ref{Fig2}\textbf{a} and \ref{Fig2}\textbf{b} at each separate
field are well fitted (dashed lines) using equations
(\ref{EqAbsSxx}) and (\ref{EqThetaSxy}) and the classical Drude
formulas:

\begin{eqnarray}
\sigma_{xx}(\omega,B) & = &\frac{2D}{\pi} \cdot
\frac{1/\tau-i\omega}{\omega_c^2-(\omega+i/\tau)^2} \label{EqSxx}\\
\sigma_{xy}(\omega,B) & = & -\frac{2D}{\pi} \cdot
\frac{\omega_c}{\omega_c^2-(\omega+i/\tau)^2} \label{EqSxy}
\end{eqnarray}

\noindent where $D$ is the Drude weight, $\omega_{c}$ is the
cyclotron frequency (which is positive for electrons and
negative for holes) and $\tau$ is the scattering time. From
here we see that the observed large Faraday rotation is
associated with the cyclotron resonance. Furthermore, the curve
$\theta(\omega)$ reveals the value and sign of the cyclotron
frequency, since $|\omega_{c}|$ coincides with the position of
the maximum absolute slope $|d\theta(\omega)/d\omega|$ (shown
by the arrow in Figure \ref{Fig2}\textbf{a} for 7 T), while the
sign of the slope matches the sign of $\omega_c$. In our case,
the negative slope is an unmistakable signature of hole doping.

In the picture of non-interacting Dirac fermions, the Drude
weight (at zero magnetic field) is equal to
$2\sigma_0|\epsilon_F|/\hbar$. Using this relation, we estimate
$|\epsilon_F| = 0.35\pm 0.03$ eV from the Drude weight
extracted from the fit. This agrees with the estimate based on
the mid-infrared absorption threshold but is subject to
relatively large error bars.

The cyclotron frequency (Fig.\ref{Fig2}\textbf{e}) demonstrates
an approximately linear growth with field, in agreement with
the theoretical relation \cite{AndoJPSJ02,GusyninNJP09}

\begin{equation}
\omega_c = \frac{e B v_{F}^{2}}{\epsilon_F}\label{EqWc}.
\end{equation}

\noindent However, the theoretical curve using the value
$\epsilon_{F}$ = -0.34 eV and $v_{F} = 1.02\times 10^6$ m/s
extracted from the mid-infrared optical spectra and ARPES data
respectively (dashed line in Fig.\ref{Fig2}\textbf{e}) is
somewhat lower than the experiment. Although the reason of this
deviation is currently unknown, one can speculate about its
possible relation to many-body effects. Interestingly an
electron-plasmon coupling was recently observed in similar
samples \cite{BostwickScience10}.

The carrier mobility $\mu$ can also be extracted using the
relation $|\omega_c| \tau = \mu |B|$
(Fig.\ref{Fig2}\textbf{f}). The mobility varies weakly with
field between 2,400-3,000 cm$^2$/(V$\cdot$s). Even at 7 T, $\mu
|B| \sim 1$, which means that the cyclotron resonance is
significantly damped due to disorder. This can also be seen
from the large width ($1/\tau \sim$ 10 meV) of the cyclotron
peak in the inset of Figure \ref{Fig2}\textbf{b}.

\begin{figure*}
\includegraphics[width=17cm]{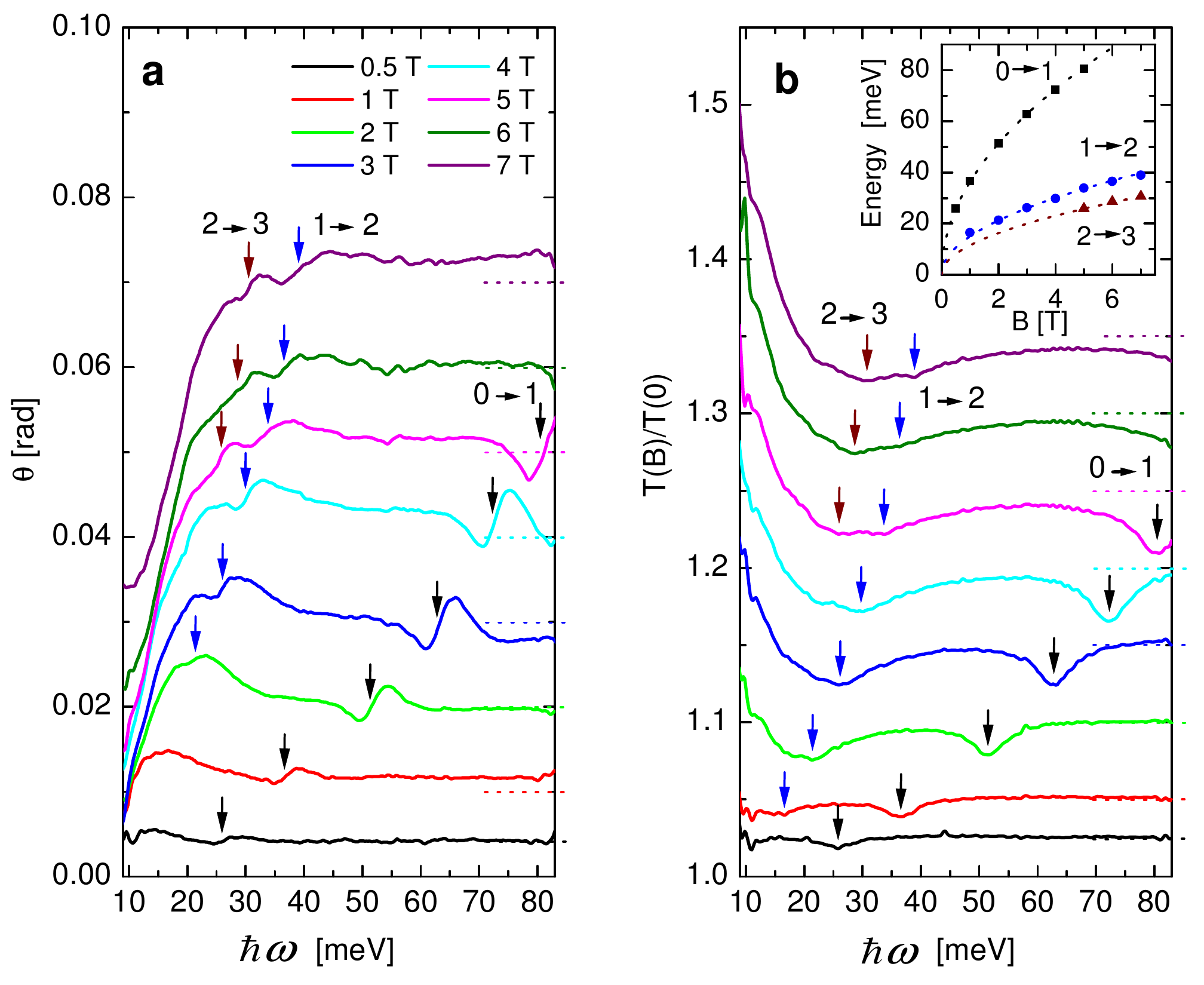}\\
\caption{\textbf{Faraday rotation and magneto-optical
transmission spectra of multi-layer graphene.} The Faraday angle
$\theta$ (panel \textbf{a}) and the transmission ratio
$T(B)/T(0)$ (panel \textbf{b}) at several fields up to 7 T at 5
K. The curves on both panels are displaced for clarity; the
offsets are given by the dashed lines of the same color. The
energies of the inter-LL transitions 0$\rightarrow$1,
1$\rightarrow$2 and 2$\rightarrow$3 are indicated by the black,
blue and dark red arrows respectively and shown as function of
magnetic field in the inset of panel \textbf{b}, where also
theoretical curves obtained using equation (\ref{EqEn}) with
$v_F = 1.00\times 10^6$ m/s are shown as dashed lines.} \label{Fig3}
\end{figure*}

The Faraday rotation and absorption observed in multi layer
graphene is strikingly different (Figure \ref{Fig3}). Both the
Faraday angle and absorption spectra show additional, strongly
field dependent resonance structures, marked by arrows. As was
found in previous studies \cite{SadowskiPRL06,JiangPRL07}, they
correspond to optical transitions between individual LLs. The
series of transitions with energies $E_{1} - E_{0}$, $E_{2} -
E_{1}$ and $E_{3} - E_{2}$ can be identified and fitted using
equation (\ref{EqEn}), which gives $v_F = 1.00\pm 0.01 \times
10^6$ m/s (inset of Fig.\ref{Fig3}\textbf{b}). Although
inter-LL absorption peaks have been observed before
\cite{SadowskiPRL06,JiangPRL07}, the measurement of the Faraday
rotation at the same conditions is the key novelty of the
present work. Now we can distinguish transitions involving
electrons from the ones involving holes. In particular, the
positive sign of $d\theta(\omega)/d\omega$ at the transition
energies, which coincide with the inflection points,
unequivocally determines that the transitions are between
electron bands. The low frequency part of the spectra features
a cyclotron-resonance like structure, similar to the data on
SLG. The sign of the Faraday angle now corresponds to electron
doping.

Since only transitions between occupied and empty levels are
allowed, the simultaneous presence of 0$\rightarrow$1,
1$\rightarrow$2 and 2$\rightarrow$3 transitions and the
low-frequency cyclotron structure in MLG arises from a
variation of the Fermi energy across different layers. The
layer closest to the substrate is highly doped
\cite{OhtaPRL07}; the doping in subsequent layers decreases
exponentially with layer number. The most strongly doped
innermost layer gives rise to the cyclotron resonance, as in
SLG. Meanwhile, the individual inter-LL transitions originate
from the weakly doped layers, which are in the quantum regime,
as can be seen from the square-root field dependence of the
transition energies. The maximum Faraday angle in this sample
is smaller than in the monolayer sample, because the innermost
layer in MLG is more weakly doped than the latter.

Summarizing the data on both samples, we point out that the
Faraday rotation is a powerful contact-free tool to distinguish
the carrier type and measure the mobility in graphene. Both in
the classical (high doping) and quantum (low doping) limits
this technique reveals magneto-optical resonances due to either
the cyclotron effect or to inter-LL transitions. The technique
has the potential to distinguish electronic contributions from
separate graphene layers. However, our most striking
observation is the giant value of the rotation ($>$ 0.1 rad for
a single atomic layer). This is much larger than the
fine-structure constant ($\sim 10^{-2}$) - the predicted scale
for the Faraday angle, associated with the quantized Hall
conductance
\cite{VolkovJETPL85,MorimotoPRL09,FialkovskyJPAMT09,IkebePRL10}.
We demonstrated that this enhancement is due to the proximity
to the resonance frequency. Although in 2DEGs the cyclotron
resonance gives rise to comparable absolute rotations
\cite{SuzukiJPSJ03,IkebePRL10}, the rotation comes from an
effective layer which is about one order of magnitude thicker
than graphene. Furthermore, while in 2DEGs the cyclotron
frequency is doping independent, in graphene it can be tuned
with doping (electrostatic or chemical) as seen from Equation
(\ref{EqWc}), and much higher frequencies can be achieved at
the same fields.

The Faraday effect and the associated magneto-optical Kerr
effect are widely used in optical communication, data storage
and computing. We suggest that the use graphene for fast
tunable ultrathin magneto-optical devices should be explored.
Indeed, the possibility of easy and fast ambipolar doping
together with a strong Faraday effect in a wide frequency range
is a unique combination not present in other known materials.

This work was supported by the Swiss National Science
Foundation (SNSF) by the grant 200021-120347, through the
National Center of Competence in Research ``Materials with
Novel Electronic Properties-MaNEP". Work in Erlangen was
supported by the German Research Council (DFG) through research
grant SE 1087/5-1 and through the Cluster of Excellence
`Engineering of Advanced Materials' at the University of
Erlangen-Nuremberg. We thank S.G. Sharapov, J. Hancock and
J.L.M. van Mechelen for discussions.


\section{\textbf{Methods}}

\subsection{Samples}

Single and multi layer epitaxial graphene were made by
graphitization of the surface of semi-insulating 6H-SiC
substrates in an argon atmosphere \cite{EmtsevNatMater09}. The
sample size was 10$\times$10 mm$^2$. SLG was obtained on the Si
terminated face of SiC by first preparing a ($6\sqrt{3}\times
6\sqrt{3}$) reconstructed surface (the so-called buffer layer)
by annealing in 1 bar of Ar at a temperature of 1450
$^{\circ}$C, which was subsequently converted into
quasi-freestanding graphene by intercalation of hydrogen
\cite{RiedlPRL09,SpeckMSF10}. MLG was grown on the C-terminated
face of SiC. In either case, the back side of the substrate was
cleaned from undesirably grown graphene using scotch-tape and
was checked to be graphene free using X-ray photo emission
(XPS). Likewise, the layer thicknesses of the SLG and MLG
samples were determined using XPS measurements
\cite{EmtsevPRB08}.

\subsection{Magneto-optical experiment}

The sample was mounted in a split-coil superconducting magnet
attached to a Fourier-transform spectrometer. A globar light
source and He-cooled bolometer detector were used. Two
grid-wire gold polarizers (one fixed and one rotating) were mounted
before and after the sample. The Faraday rotation was deduced
from the angle of the rotating polarizer at which the intensity
was at minimum. The substrate-normalized transmission $T$ was
measured with one polarizer only and is defined as the ratio
between the intensity of the light transmitted through the
graphene on SiC and through bare substrate.

In the thin-film limit, the absorption and the Faraday angle are related to
the diagonal and the Hall conductivity respectively by equations (2) and (3), where we
made use of a spectrally smooth dimensionless function
$f_s(\omega)$, which depends on the refractive index
$n_s(\omega)$ and the extinction coefficient $k_s(\omega)$ and the thickness $d$ of
the substrate. In our measurements, the Fabry-Perot interference in the substrate was not spectrally resolved.
In this case:
\begin{equation}
f_s(\omega) = \left(\frac{1}{n_s +
1}+\frac{2n_s}{n_{s}^2-1}\cdot\frac{q^2}{1-q^2}\right)
\label{EqSxxAp}
\end{equation}
\noindent where $q = \left[(n_s - 1)/(n_s + 1)\right]^2
\exp\left[-(2\omega/c) k_s d\right]$.

Note that for a weakly absorbing substrate $q \approx
\left[(n_s - 1)/(n_s + 1)\right]^2$ and therefore $f_{s}\approx
(n_s^2 + 3)/(4 n_s^2 + 4)$ while for a strongly absorbing
substrate $q \ll 1$ and $f_{s}\approx 1/(n_s + 1)$.
Experimentally, both $n_s$ and $k_s$ can be determined from the
measurement of the absolute transmission and reflection spectra
of the bare substrate.
Figure \ref{Fig4} shows the function $f_{s}$ of the used SiC substrate ($d\approx 370 $ microns) at 5 K.
Below 60 meV, this function is constant; at higher photon energies it decreases because of the
presence of a strong optically active phonon in SiC.


\begin{figure*}
\includegraphics[width=12cm]{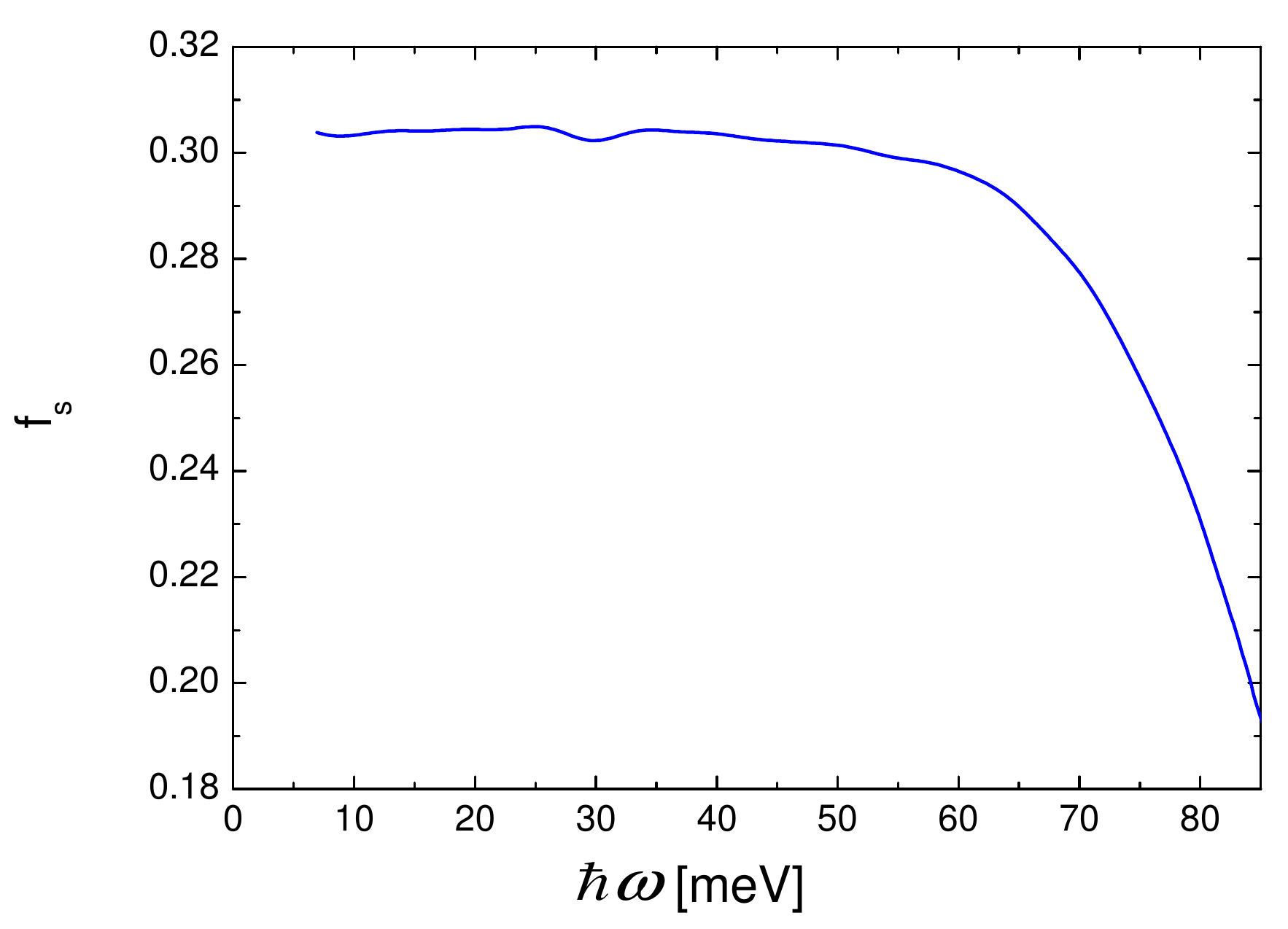}\\
\caption{The function $f_s(\omega)$ used in the formulas (\ref{EqAbsSxx}) and (\ref{EqThetaSxy}).}
\end{figure*}\label{Fig4}


\newpage




\begin{thebibliography}{99}

\bibitem{GeimNM07}
Geim, A.K. \& Novoselov K.S. The rise of graphene, \emph{Nature
Mater.} \textbf{6}, 183 (2007).

\bibitem{CastroNetoRMP09}
Castro Neto, A.H., Guinea, F., Peres, N.M.R., Novoselov K.S.,
\& Geim, A.K. The electronic properties of graphene, \emph{Rev.
Mod. Phys.} \textbf{81}, 109 (2009).

\bibitem{GusyninPRL05} Gusynin, V. P. \& Sharapov S. G. Uniconventional Integer Quantum Hall Effect in Graphene,\emph{ Phys. Rev.
Lett.} \textbf{95,} 146801 (2005).

\bibitem{PeresPRB06} Peres, N. M. R., Guinea, F. \& Castro Neto, A. H. Electronic properties of disordered two-dimensional carbon,\emph{ Phys. Rev.
B} \textbf{73,} 125411 (2006).

\bibitem{NovoselovNat05}Novoselov, K. S. \emph{et al.}, Two--dimensional gas of massless Dirac fermions in graphene, \emph{Nature} \textbf{438},
197-200 (2005). 

\bibitem{ZhangNat05} Zhang, Y., Tan, Y-W., Stormer, H. L. \& Kim, P. Experimental observation of the quantum Hall effect and Berry's phase in
graphene,
\emph{Nature} \textbf{438}, 201-204 (2005). 

\bibitem{JobstPRB10}Jobst J. \emph{et al.} Quantum oscillations and quantum Hall effect in epitaxial graphene
\emph{ Phys. Rev. B} \textbf{81}, 195434 (2010).

\bibitem{ShenAPL09}Shen, T. \emph{et al.} Observation of quantum-Hall effect in gated epitaxial graphene grown on SiC (0001)  \emph{Appl. Phys. Lett.}
\textbf{95}, 172105 (2009).

\bibitem{WuAPL09} Wu, X. \emph{et al.} Half integer quantum Hall effect in high mobility single layer epitaxial graphene  \emph{Appl. Phys. Lett.}
\textbf{95}, 223108 (2009).


\bibitem{AndoJPSJ02} Ando, T., Zheng, Y. \& Suzuura, H., Dynamical Conductivity and Zero-Mode Anomaly in Honeycomb Lattices, \emph{J.
Phys. Soc. Jpn.} \textbf{71}, 1318–1324 (2002).


\bibitem{FalkovskyEPJ07} Falkovsky, L.A.\& Varlamov, A.A. Space-time dispersion of graphene
conductivity, \emph{Eur. Phys. J. B} \textbf{56}, 281-284
(2007).

\bibitem{KuzmenkoPRL08} Kuzmenko, A. B., van Heumen, E., Carbone, F. \& van der Marel, D. Universal optical conductance of graphite,\emph{ Phys. Rev.
Lett.} \textbf{100,} 117401 (2008).

\bibitem{NairSc08} Nair, R.R. \emph{et al.} Fine structure constant defines visual transparancy of graphene, \emph{Science} \textbf{320}, 1308 (2008).

\bibitem{LiNaturePhys08} Li, Z.Q. \emph{et al}. Dirac charge dynamics in graphene by infrared
spectroscopy, \emph{Nature Phys.} \textbf{4}, 532 (2008).

\bibitem{GusyninNJP09} Gusynin, V. P., Sharapov, S. G. \& Carbotte, J. P. On the universal AC optical background in
graphene, \emph{New J. Phys.} \textbf{11}, 095013 (2009).

\bibitem{SadowskiPRL06} Sadowski, M.L., Martinez, G., Potemski, M., Berger, C. \& de
Heer, W.A. Landau Level Spectroscopy of Ultrathin Graphite
Layers, \emph{Phys. Rev. Lett.} \textbf{97}, 266405 (2006).

\bibitem{JiangPRL07} Jiang, Z. \emph{et al}. Infrared spectroscopy of Landau levels in graphene, \emph{Phys. Rev. Lett.} \textbf{98}, 197403 (2007).

\bibitem{GusyninJPCM07} Gusynin, V. P., Sharapov, S. G. \& Carbotte, J. P., Magneto-optical conductivity in Graphene,
\emph{J. Phys. Condens. Matter} \textbf{19}, 026222 (2007).

\bibitem{OrlitaPRL08} Orlita, M. \emph{et al.} Approaching the
dirac point in high-mobility multilayer epitaxial graphene.
\emph{Phys. Rev. Lett.} \textbf{101}, 267601 (2008).

\bibitem{KaplanPRL96} Kaplan, S.G. \emph{et al}. Normal state ac Hall effects in YBa2Cu3O7 thin
films, \emph{Phys. Rev. Lett.} \textbf{76}, 696-699 (1996).

\bibitem{MorimotoPRL09} Morimoto, T., Hatsugai, Y. \& Aoki, H. Optical Hall Conductivity in Ordinary and Graphene Quantum Hall Systems, \emph{Phys.
Rev. Lett.}
\textbf{103}, 116803 (2009). 

\bibitem{FialkovskyJPAMT09} Fialkovsky, I.V. \& Vassilevich,
D.V., Parity-odd effects and polarization rotation in graphene,
\emph{J. Phys. A: Math. Theor.} \textbf{42} 442001 (2009).

\bibitem{BergerJPCB04} Berger, C. \emph{et al.}, Ultrathin
epitaxial graphite: 2d electron gas properties and a route
toward graphene-based nanoelectronics, \emph{J. Phys. Chem. B},
\textbf{108}, 19912 (2004).

\bibitem{EmtsevNatMater09} Emtsev, K.V. \emph{et al.}, Towards wafer-size graphene layers by atmospheric pressure graphitization of silicon carbide,
\emph{Nature Mater.} \textbf{8}, 203 (2009).

\bibitem{RiedlPRL09} Riedl, C., Coletti C., Iwasaki T., Zakharov, A.A. \&
Starke, U. Quasi-Free-Standing Epitaxial Graphene on SiC
Obtained by Hydrogen Intercalation, \emph{Phys. Rev. Lett.}
\textbf{103}, 246804 (2009).

\bibitem{SpeckMSF10} Speck, F. \emph{et al.} Quasi-freestanding Graphene on SiC(0001), \emph{Mat. Sci.
Forum}, \textbf{645-648}, 629-632 (2010).

\bibitem{BostwickScience10} Bostwick, A. \emph{et al}.
Observation of plasmarons in quasi-free-standing doped
graphene, \emph{Science}, \textbf{328}, 999 (2010).

\bibitem{OhtaPRL07} Ohta, T. \emph{et al.} Interlayer Interaction and Electronic Screening in Multilayer Graphene Investigated with Angle--Resolved
Photoemission Spectroscopy, \emph{Phys. Rev. Lett.}
\textbf{98}, 206802 - 206804 (2007).

\bibitem{VolkovJETPL85} Volkov, V.A. \& Mikhailov, S.A., Quantization Of The Faraday-Effect In Systems With A Quantum
Hall-Effect, \emph{JETP Lett.} \textbf{41}, 476-478 (1985).

\bibitem{IkebePRL10}Ikebe, Y. \emph{et al.}, Optical Hall Effect in the Integer Quantum Hall regime \emph{Phys. Rev. Lett.}
\textbf{104}, 256802 (2010).

\bibitem{SuzukiJPSJ03}Suzuki, M., Fujii, K., Ohyama, T., Kobori, H. \& Kotera, N. Far-Infrared Resonant Faraday Effect in Semiconductors, \emph{J.
Phys. Soc. Jpn.} \textbf{72}, 3276-3285 (2003). 

\bibitem{EmtsevPRB08} Emtsev, K. V., Speck, F., Seyller, Th., Ley, L.\& Riley, J. D.
Interaction, growth, and ordering of epitaxial graphene on
SiC\{0001\} surfaces: A comparative photoelectron spectroscopy
study, \emph{Phys. Rev. B} \textbf{77}, 155303 (2008).

\end{thebibliography}
\end{document}